\documentstyle[prl,aps]{revtex}
\begin{document}
\draft
\title{Numerical evaluation of a two loop diagram in the cut-off
regularization}
\author{Ji-Feng Yang, Jie Zhou\footnote{Graduate student.} and
Chen Wu\footnote{Graduate student. }}
\address{Department of Physics, ECNU, Shanghai, 200062, China}

%\date{}
\maketitle

\begin{abstract}
The sunset diagram of $\lambda\phi^4$ theory is evaluated
numerically in cutoff scheme and a nonzero finite term (in
accordance with dimensional regularization (DR) result) is found
in contrast to published calculations. This finding dramatically
reduces the critical couplings for symmetry breaking in the two
loop effective potential discussed in our previous work.
\end{abstract}

\section{Introduction}
Dimensional regularization\cite{DR} has become an efficient
regularization scheme for theoretical calculations in high energy
physics. It makes analytical results more available, and its gauge
invariance almost decisively defeated the other schemes (say
cutoff scheme) in QCD and other gauge theories.

However, things are not so simple. It is very natural to
anticipate that DR might fail sooner or later somewhere as it is
just a special regularization but never a true theory of the short
distance physics. Such event is now known: the $\overline{MS}$
scheme does not provide a physical prediction for low energy
nucleons scattering in case of long scattering lengths\cite{KSW}
in nonperturbative context. To yield meaningful results,
unconventional subtraction is devised and now known as PDS
\cite{KSW}\footnote{There are other schemes that could yield
similar predictions as PDS did, see, e.g., Ref.\cite{mehen}.}.
Since this is obtained with hindsight, we should not discard the
cutoff scheme without further exploiting its virtues. There have
recently been some further investigations over the relations
between the divergences in DR and cutoff schemes and the
associated limitations\cite{DRCutoff,ComTP}. The conventional
virtues of the DR scheme were shown to be due to an implicit
subtraction\cite{DRCutoff}. In nonperturbative contexts involving
multiloop corrections the DR might even make renormalization
infeasible\cite{ComTP}.

Traditionally, it is very hard to obtain analytical results for
multiloop amplitudes within the cutoff scheme. In this report we
show that one could achieve this goal numerically at a pretty good
precision. We will make use of the dynamical symmetry breaking of
$\lambda\phi^4$ discussed in \cite{prd} to illustrate the
numerical evaluation of the finite constant term in the two loop
sunset diagram. Thus the report is organized as follows. First we
quote some results of Ref.\cite{prd} in Sec. II as preparation. In
Sec. III we present our strategy and procedures for doing the
numerical analysis of the problem and its possible prospective
significance. The result is also given there. In Sec. IV we
revisit the problem introduced in Sec. II to show the significance
of the new result obtained in Sec. III. The summary will also be
given there.

\section{renormalization of the two loop effective
potential} In\cite{prd}, the dynamical symmetry breaking of
massless $\lambda\phi^4$ model with $Z_2$ symmetry was studied in
two loop effective potential. The algorithm in use is well known
according to Jackiw\cite{Jackiw},
\begin{eqnarray}
&&V_{(2l)}\equiv \lambda\phi^4+\frac{1}{2}I_{0}(\Omega)+3\lambda
I^2_1(\Omega)-48\lambda^2 \phi^2I_2(\Omega),\\
&&\Omega\equiv \sqrt{12\lambda\phi^2};\ \ \
I_{0}(\Omega)=\int\frac{d^4k}{(2\pi)^4}\ln(1+\frac{\Omega^2}{k^2});
\ \ \ I_1(\Omega)=\int\frac{d^4 k}{(2\pi)^4}\frac{1}{k^2+\Omega^2};\\
\label{I} &&I_2(\Omega)=\int\frac{d^4 k d^4
l}{(2\pi)^8}\frac{1}{(k^2+\Omega^2)(l^2+\Omega^2)((k+l)^2
+\Omega^2)}.
\end{eqnarray} Here we have Wick rotated all the loop integrals
into Euclidean space. The loop integrals and the effective
potential had been calculated in literature \cite{Jackiw,FJ}. For
comparison, we list out below only the results for the sunset
diagram defined by $I_2 (\Omega)$ in Eq.(~\ref{I}) , the only two
loop diagram with overlapping divergences.
\subsection{Loop amplitudes in DR and cutoff schemes}From
\cite{FJ} we find that
\begin{eqnarray}
\mu^{4\epsilon}I^{(D)}_2(\Omega)=-\frac{3\Omega^2}{2(4\pi)^4
\epsilon^2}\{1+(3-2\overline{L}) \epsilon+
(2\overline{L}^2-6\overline{L}+7+6S-\frac{5}{3}\zeta(2)
)\epsilon^2\}
\end{eqnarray} with $S=\sum^{\infty}_{n=0}\frac{1}{(2+3n)^2},
\overline{L}=L+\gamma-\ln 4\pi$ and $L=\ln
\frac{\Omega^2}{\mu^2}$. While from \cite{Jackiw},
\begin{eqnarray}
\label{cutoff}
I^{(\Lambda)}_{2}(\Omega)=\frac{1}{(4\pi)^4}\{2\Lambda^2
-\frac{3\Omega^2}{2}\ln^2 \frac{\Omega^2}{\Lambda^2}+3\Omega^2\ln
\frac{\Omega^2}{\Lambda^2}+o(\Lambda^{-2})\}.
\end{eqnarray}Note that the finite double logarithmic term
(leading term) disagree between the two regularization schemes,
the disagreement is removed only after all subdivergences are
removed\cite{ComTP}.

We note in particular that the term proportional to $\Omega^2$ in
the two loop integral is not explicitly given in \cite{Jackiw},
and it is this finite term and its numerical estimation that is
our main concern in this paper.

\subsection{Prescription dependence}
The renormalization were performed in Ref.\cite{prd} in several
prescriptions, $\overline{MS}$ for $V^{(D) }_{(2l)}( \Omega )$
\cite{FJ}, Jackiw\cite{Jackiw}, Coleman-Weinberg (CW)\cite{CW} and
$\mu^2_{\Lambda}$ (defined in\cite{prd}, Appendix B) for
$V^{(\Lambda )}_{(2l)}(\Omega )$ with the renormalized potential
taking the following form
\begin{eqnarray}
\label{SCH} V_{(2l)}( \Omega )
=\Omega^{4}\{\frac{1}{144\lambda}+\frac{L-1/2}{(8\pi)^2}
+\frac{3\lambda}{(4\pi)^4}[ L^2+2(L-1)^2+\alpha]\},
\end{eqnarray}where $L=\overline{L}$, $\ln
\frac{\Omega^2}{\mu^2_{\Lambda}}$, $\ln
\frac{\Omega^2}{12\lambda\mu^2_{Jackiw}}$ and $\ln
\frac{\Omega^2}{12\lambda\mu^2_{CW}}-\frac{25}{6}$ in respectively
$\overline{MS}$, $\mu^2_{\Lambda}$, Jackiw and Coleman-Weinberg
schemes. In all the above formulas the scheme dependence of field
strength and coupling constant are understood. The explicit
intermediate renormalization prescriptions dependence of the
effective potential expressed by $\alpha$ is listed in
Table~\ref{table1}.

The dynamical symmetry breaking solution in the two loop effective
potential defined in Eq.~(\ref{SCH}) could be obtained from the
first order condition $ \frac{\partial
V_{(2l)}(\Omega(\phi))}{\partial\phi}=0$ and the solutions
read\cite{prd},
\begin{eqnarray}
\phi^2_{+}(\lambda;[\mu,\alpha])=\frac{\mu^2}{12\lambda} \exp
\left \{ \frac{1}{6} \left [ 1-\frac{4\pi^2}{3\lambda}+
\sqrt{\frac{1}{3}[4-36\alpha-(1+\frac{4\pi^2}{\lambda})^2]} \right
] \right\},
\end{eqnarray}whose existence requires that
\begin{eqnarray}
\label{alpha} \alpha < \frac{1}{12},\ \ \lambda \geq
\lambda_{cr}\equiv\frac{4\pi^2}{\sqrt{4-36\alpha}-1}.
\end{eqnarray}That means not all schemes are consistent with
symmetry breaking. In addition, for the symmetry breaking
solutions to be stable, the coupling must be further
constrained\cite{prd},
\begin{eqnarray}
\label{critical} \lambda\geq \hat{\lambda}_{cr}\equiv
\frac{4\pi^2}{\sqrt{4-36\alpha-27}-1},\ (>\lambda_{cr}).
\end{eqnarray}Both Eq.~(\ref{alpha}) and ~(\ref{critical}) are
summarized in Table~\ref{table2}.

If the finite term proportional to $\Omega^2$ in $I_2(\Omega)$
existed, the $\alpha$ in $\mu^2_{\Lambda}$ and Jackiw
prescriptions would consequently be replaced by $-2-\frac{4C}{3}$
and $-\frac{5}{4}-\frac{4C}{3}$ in the following parameterization
of Eq.~(\ref{cutoff})\footnote{In Coleman-Weinberg prescription
the definition of $\alpha=\frac{49}{3}$ is not altered due to its
special renormalization condition\cite{Jackiw,CW}. },
\begin{eqnarray}
I^{(\Lambda)}_{2}(\Omega)=\frac{1}{(4\pi)^4}\{2\Lambda^2
-\frac{3\Omega^2}{2}\ln^2 \frac{\Omega^2}{\Lambda^2}+3\Omega^2\ln
\frac{\Omega^2}{\Lambda^2}+C\Omega^2+o(\Lambda^{-2})\}.
\end{eqnarray}It is not difficult to see that a different value of
$\alpha$ means different symmetry breaking status, hence different
physics, a nonperturbative scheme dependence problem as emphasized
in ref.\cite{prd}. It is therefore important to determine the
value of $C$ in cutoff scheme in order to be sure that the Jackiw
and $\mu^2$ prescription are really consistent symmetry breaking
in two loop effective potential. Thus the rest of this report will
be devoted to estimate this constant term numerically.
\section{Strategy for numerical estimation of $c$}
First let us put the sunset diagram into the following form after
integrating out the angular variables,
\begin{eqnarray}
\label{I2} I_2(M)=\frac{1}{2(4\pi)^4}\int^{\beta}_0
\frac{dx}{x+M}\int^{\beta}_0 \frac{dy}{y+M}\left
\{x+y+M-\sqrt{(x+y+M)^2-4xy}\right \}
\end{eqnarray}with $M\equiv\Omega^2, x\equiv k^2, y\equiv l^2$ and
$\beta\equiv\Lambda^2$. The asymptotic form of this integral as
cutoff tends to infinity reads
\begin{eqnarray}
\label{OO} I_2(M)&=&\frac{1}{(4\pi)^4}\left
\{2\beta-\frac{3M}{2}\ln^2\frac{M}{\beta}+3M\ln\frac{M}{\beta}+CM
\right \}+o(M/\beta) \nonumber \\
&\equiv&I_2^{(asy)}(M)+\frac{CM}{(4\pi)^4}+o(M/\beta).
\end{eqnarray}In order to determine the constant $C$ we compare
$I_2(M)$ and $I_2^{(asy)}(M)$ after we evaluated the former
numerically.

The feasibility of numerical treatment of an apparently divergent
integral is guaranteed by the observation that the cutoff need not
be very large. Eq.(~\ref{OO}) tells us that the analytical source
of error comes from $\frac{M}{\beta}$. The magnitude of the
constant $C$ should be of order 1 or 10 in the limit
$\Lambda^2\rightarrow\infty$. Thus taking $\frac{M}{\beta}$ to be
$10^{-4}$ or $10^{-5}$ should be sufficient for our purpose. There
is a main obstacle in the practice for numerical estimation of
divergent integral in the cutoff scheme: the presence of power law
pieces, which consume most of the computer's capacity in order to
get a finite number at a precision of $10^{-4}$. The larger the
cutoff is chosen, the less precision capacity is left over for
finite numbers. We must get rid of such pieces. The remaining
logarithmic terms save most of the computer capacity for the
finite numbers, as even taking the cutoff as large as $10^{10}M$
will only yield a double log term at order $\sim 100\ln 10\sim
230$, which consumes almost no precision capacity. This goal could
be readily achieved at least in two ways: for the sunset diagram,
the power law term $\sim 2\beta$ could be removed either by
subtracting it explicitly in the integrand or by differentiating
the whole integral with respect to $M$. In fact we will report
both approaches' result.

There is another subtle point that need careful consideration, the
step length for the numerical integration. Taking the step too
small will improve the precision analytically but cost a lot of
time and might accumulate numerical errors, especially in personal
computer. Usually a divergent integral has a smooth integrand
except in a finite range where the integrand is steeply changing.
The step must be small enough within such steep regions. The last
concern is to spare both the computer time and numerical error we
should perform the analytical calculation as far as we can. Thus
we will first perform one integral analytically (this is feasible
in most cases) while leave the remaining one to computer.

In the following we will describe one approach of our numerical
treatment in detail, the one with differentiation with respect to
$M$. That is, to extract $C$ we just consider the following double
integral which at most possesses logarithmic divergence,
\begin{eqnarray}
\label{pI2} &&\frac{\partial I_2(M)}{\partial
M}=-\frac{1}{(4\pi)^4}\left \{I_1-I_2+\frac{1}{2}(I_3-I_4)\right
\}= \frac{\partial I_2^{(asy)}(M)}{\partial
M}+\frac{C}{(4\pi)^4}+o(M/\beta),
\\&&I_1\equiv \int^\beta_0\frac{dx}{(x+M)^2}\int^\beta_0
\frac{dy}{y+M}(x+y+M)\ ,\\
&&I_2\equiv\int^\beta_0\frac{dx}{(x+M)^2}\int^\beta_0
\frac{dy}{y+M}\sqrt{(x+y+M)^2-4xy}\ ,\\
&&I_3\equiv\int^\beta_0\frac{dx}{x+M}\int^\beta_0
\frac{dy}{y+M} \ ,\\
&&I_4\equiv \int^\beta_0\frac{dx}{x+M}\int^\beta_0
\frac{dy}{y+M}\frac{x+y+M}{\sqrt{(x+y+M)^2-4xy}}\ .
\end{eqnarray}It is not difficult to see that the $2\beta$ piece
is removed in all the four integrals $I_1, I_2, I_3$ and $I_4$
there is at most double log terms. Then it is also easy to perform
two folds integrals once. Moreover, $I_1$ and $I_3$ can be done
analytically at all. Then the results read
\begin{eqnarray}
&&I_1=\ln\frac{\beta+M}{M}\left
\{\ln\frac{\beta+M}{M}+\frac{M}{\beta+M}-1 \right \}
+\frac{\beta^2}{M(M+\beta)}\ , \\&& I_2=\int^\beta_0
\frac{dx}{x+M} \left \{-\frac{\sqrt{R}}{Y}+\frac{x}{\sqrt{a}}
\sinh^{-1}\frac{a-xY}{2Y\sqrt{xM}}
+\sinh^{-1}\frac{Y-x}{2\sqrt{xM}}\right \}\|^{Y=\beta+M}_{Y=M}\
,\\&&I_3=\ln^2 \frac{\beta+M}{M}\ ,\\&&I_4= \int^\beta_0
\frac{dx}{x+M}\left \{
\frac{-x}{\sqrt{a}}\sinh^{-1}\frac{a-xY}{2Y\sqrt{xM}}
+\sinh^{-1}\frac{Y-x}{2\sqrt{xM}}\right \} \|^{Y=\beta+M}_{Y=M}
\end{eqnarray}with $R\equiv a-2xY+Y^2,\ \ a\equiv x^2+4xM$. There
is a seemingly singularity in the integrands ($ \sim
\frac{1}{\sqrt{xM}}$ as $x\sim0$) in both $I_2$ and $I_4$ which
does not materialize after one completes the operations on the
variable $Y$,
\begin{eqnarray}
\label{I2} && I_2=\int^\beta_0 \frac{dx}{x+M} \left
\{-\frac{\sqrt{R}}{Y}\|^{Y=\beta+M}_{Y=M}
+\frac{\sqrt{x}}{\sqrt{x+4M}}
\ln\frac{u+\sqrt{u^2+1}}{d+\sqrt{d^2+1}}
+\ln\frac{s+\sqrt{s^2+1}}{t+\sqrt{t^2+1}}\right \}\ ,\\
\label{I4}&&I_4= \int^\beta_0 \frac{dx}{x+M}\left \{
\frac{-\sqrt{x}}{\sqrt{x+4M}}\ln\frac{u+\sqrt{u^2+1}}{d+\sqrt{d^2+1}}
+\ln\frac{s+\sqrt{s^2+1}}{t+\sqrt{t^2+1}}\right \}\ ,\\&&u \equiv
\frac{x(x+3M-\beta)}{2(\beta+M)\sqrt{xM}},\ \ v\equiv
\frac{x^2+3xM}{2M\sqrt{xM}},\ \ s\equiv
\frac{\beta+M-x}{2\sqrt{xM}},\ \ t\equiv \frac{M-x}{2\sqrt{xM}}.
\end{eqnarray}

The only necessary numerical works to be done are the two one fold
integrals defined in Eq.(~\ref{I2},~\ref{I4}). As the integrand
become steeper as $x$ is closer to the origin, thus to reduce the
error from numerical integration we separate the integral into
several intervals in order to match between the step size and the
steepness of the integrand. Taking $M$ to be 1 will make the
numerical integration easy and the $C$ has been estimated with two
cutoff scale, $\beta/M\equiv\Lambda^2/\Omega^2=10^4,\ 10^5$. The
match between the interval and step size is shown in
Table~\ref{table3} and ~\ref{table4} respectively for
$\beta/M=10^4$ and $\beta/M=10^5$.

Subtracting $\frac{\partial I_2^{(asy)}(M)}{\partial M}$ from the
numerically obtained integral defined in Eq.(~\ref{pI2}) we find
the constant $C$: $C=4.13048$ for $\beta/M=10^4$ and $C=4.15412$
for $\beta/M=10^5$. The interval $(0, 10^{-6}]$ is not included as
its contribution could at most be of order $10^{-4}$. Noting that
one cutoff is order greater than the other, the agreement of the
two cases is striking. Moreover, if one try another approach,
i.e., by directly subtracting $I_2$'s integrand a term that upon
integration will yield the $2\beta$ piece, one could also obtain
an estimate of $C$ with the results read: for $\beta/M=10^4$,
$C=4.14128$ with a homogeneous step $10^{-4}$, $C=4.14274$ with a
homogeneous step $10^{-3}$, $C=4.156178$ with a homogeneous step
$10^{-2}$; for $\beta/M=10^5$, $C=4.160739$ with a homogeneous
step $10^{-2}$, $C=4.1633$ with a homogeneous step $0.005$; for
$\beta/M=10^6$, $C=4.162993$ with a homogeneous step $10^{-2}$.
Combining the two approaches' results we could safely conclude
that $C$ is not zero. To be more conservative we anticipate that
the true value should lie in the interval $[4, 4.3]$.

Before ending this section, we should point out and stress that in
order to evaluate divergent multiloop integrals numerically in
cutoff regularization: (i) one does not need a very large cutoff
scale, (ii) the power law term can be easily removed in order to
save capacity for more important task (like determination of the
finite parts). The numerical workload could further be reduced if
one varies the step size with steepness of the integrand.
\section{New predictions and summary}
With the new constant we could reevaluate the critical couplings
for symmetry breaking with the result summarized in Table~\ref{T5}
and ~\ref{T6}. We take $C$ to be 4 when calculating the critical
couplings. It is not difficult to see that the critical couplings
in $\mu^2_\Lambda$ and Jackiw prescriptions are dramatically
lowered, or the symmetry breaking might take place at much lower
values of the coupling. The interesting thing is, the
$\overline{MS}$ critical coupling, which is smaller than the ones
in the prescriptions based on old calculation of the sunset
diagram \cite{Jackiw}, now becomes a larger one than the obtained
with the new numerically determined constant $C$, and the larger
the $C$ is, the smaller the critical couplings are. In this sense,
the cutoff regularization is preferable to DR, just like the case
in the EFT applications in nucleon interactions \cite{Epel}.

Generally one would expect the critical coupling to be close to 1
instead of being very large (like 10.7) for symmetry breaking to
take place. Therefore it is a pleasing finding that constant $C$
is larger than 0 and is no less than 4. If $C=20$, a reasonable
value, the critical coupling could be 1.5238 and 1.5542. Of
course, as the potential is only calculated at two loop level, the
precise value of the critical coupling could not be taken very
serious. However, one should still be pleased to see that the
critical couplings are dramatically lowered after more careful
evaluation in the regularization scheme that is widely held as
inferior to DR, even though the computation is not exact one.

In summary, we reevaluated the sunset diagram in $\lambda\phi^4$
in the cutoff regularization half analytically and half
numerically and found that the finite local constant, which is
usually taken to be zero, is not zero and this constant could
dramatically lower the critical couplings for dynamical symmetry
breaking in the two loop effective potential in a number of
renormalization prescriptions. The important byproduct is that the
numerical calculation could be efficiently done in cutoff
regularization for divergent multiloop diagrams and we have
illustrated that the cutoff scale need not be too large.

\begin{table}[t]
\caption{$\alpha$ in various schemes}
\begin{center}
\begin{tabular}{c|c}  &\\[-.4cm]
 Scheme &  $\alpha$\\ \hline &\\[-.4cm]
$\overline{MS}$& $-2.6878$\\
$\mu^2_{\Lambda}$& $-2$\\
Jackiw & $-\frac{5}{4} $\\
Coleman-Weinberg & $\frac{49}{3} $\\
\end{tabular}
\end{center}
\label{table1}
\end{table}

\begin{table}[t]
\caption{Critical values of $\lambda$}
\begin{center}
\begin{tabular}{c|c|c}  &&\\[-.4cm]
 Scheme & $\lambda_{cr}$&$\hat{\lambda}_{cr}$\\ \hline &&\\[-.4cm]
$\overline{MS}$&  4.368& $5.2024$\\
$\mu^2_{\Lambda}$& 5.1152 & 6.5797 \\
Jackiw & 6.5797 & 10.698\\
\end{tabular}
\end{center}
\label{table2}
\end{table}

\begin{table}[t]
\caption{Match I for $\beta/M=10^4$}
\begin{center}
\begin{tabular}{c|c|c|c|c|c|c|c}  &&\\[-.4cm]
Interval & $[10^{-6}, 10^{-2}]$ & $[10^{-2}, 10^{-1}]$ &
$[10^{-1}, 10^0]$ & $ [10^0,
10^1]$ & $[10^1, 10^2]$ & $[10^2, 10^3]$ & $[10^3, 10^4]$ \\
\hline &&
\\[-.4cm]Step Size & $10^{-8}$ & $10^{-7}$ & $10^{-6}$ & $10^{-5}$ &
$10^{-4}$ & $10^{-3}$ & $ 10^{-3}$\\
\end{tabular}
\end{center}
\label{table3}
\end{table}

\begin{table}[t]
\caption{Match II for $\beta/M=10^5$}
\begin{center}
\begin{tabular}{c|c|c|c|c|c|c|c|c} &&\\[-.4cm]
Interval & $[10^{-6}, 10^{-2}]$ & $[10^{-2}, 10^{-1}]$ &
$[10^{-1}, 10^0]$&$[10^0, 10^1]$ & $[10^1, 10^2]$ & $[10^2, 10^3]$
& $[10^3, 10^4]$ & $[10^4, 10^5]$\\ \hline &&\\[-.4cm]Step Size &
$10^{-8}$ & $10^{-7}$ & $10^{-6}$ & $10^{-5}$ & $10^{-4}$ & $10^{-3}$
 & $ 10^{-3}$&$10^{-3}$\\
\end{tabular}
\end{center}
\label{table4}
\end{table}

\begin{table}[t]
\caption{ New $\alpha$'s}
\begin{center}
\begin{tabular}{c|c}  &\\[-.4cm]
 Scheme &  $\alpha$\\ \hline &\\[-.4cm]
$\overline{MS}$& $-2.6878$\\
$\mu^2_{\Lambda}$& $-7\frac{1}{3}$\\
Jackiw & $-6\frac{7}{12} $\\
Coleman-Weinberg & $16\frac{1}{3} $\\
\end{tabular}
\end{center}
\label{T5}
\end{table}

\begin{table}[t]
\caption{New critical values of $\lambda$}
\begin{center}
\begin{tabular}{c|c|c}  &&\\[-.4cm]
 Scheme & $\lambda_{cr}$&$\hat{\lambda}_{cr}$\\ \hline &&\\[-.4cm]
$\overline{MS}$&  4.368& $5.2024$\\
$\mu^2_{\Lambda}$& 2.5684& 2.7181 \\
Jackiw & 2.7181 & 2.8967\\
\end{tabular}
\end{center}
\label{T6}
\end{table}
\section*{Acknowledgement}
Ji-Feng Yang thanks W. Zhu for his helpful discussions. This work
is supported in part by the National Natural Science Foundation of
China under Grant No.s 10075020 and 10205004.

\end{document}